# Real-Space Imaging of the Tailored Plasmons in Twisted Bilayer Graphene

F. Hu[1,2]*, Suprem R. Das[2,3,4,5]*, Y. Luan[1,2]*, T.-F. Chung[6,7], Y. P. Chen[6,7,8,9], Z. Fei[1,2]†

[1]Department of Physics and Astronomy, Iowa State University, Ames, Iowa 50011, USA
[2]Ames Laboratory, U.S. Department of Energy, Iowa State University, Ames, Iowa 50011, USA
[3]Department of Mechanical Engineering, Iowa State University, Ames, Iowa 50011, USA
[4]Department of Industrial and Manufacturing Systems Engineering, Kansas State University, Manhattan, KS 66506, USA
[5]Department of Electrical and Computer Engineering, Kansas State University, Manhattan, KS 66506, USA
[6]Birck Nanotechnology Center, Purdue University, West Lafayette, Indiana 47907, USA
[7]Department of Physics and Astronomy, Purdue University, West Lafayette, Indiana 47907, USA
[8]School of Electrical and Computer Engineering, Purdue University, West Lafayette, Indiana 47907, USA
[9]Purdue Quantum Center, Purdue University, West Lafayette, Indiana 47907, USA

*These authors contributed equally to this work.

†Corresponding author: Z.F. (zfei@iastate.edu)

**Abstract**

We report a systematic plasmonic study of twisted bilayer graphene (TBLG) – two graphene layers stacked with a twist angle. Through real-space nanoimaging of TBLG single crystals with a wide distribution of twist angles, we find that TBLG supports confined infrared plasmons that are sensitively dependent on the twist angle. At small twist angles, TBLG has a plasmon wavelength comparable to that of single-layer graphene (SLG). At larger twist angles, the plasmon wavelength of TBLG increases significantly with apparently lower damping. Further analysis and modeling indicate that the observed twist-angle-dependence of TBLG plasmons in the Dirac linear regime is mainly due to the Fermi-velocity renormalization, a direct consequence of interlayer electronic coupling. Our work unveils the tailored plasmonic characteristics of TBLG and deepens our understanding of the intriguing nano-optical physics in novel van der Waals (vdW) coupled two-dimensional (2D) materials.

**Main text**

Graphene Dirac plasmons [1-6], which are collective oscillations of Dirac fermions in graphene, have been widely investigated in recent years by using both the electron energy loss spectroscopy [7-9] and optical imaging/spectroscopy [10-21] techniques. These quasiparticles demonstrate many superior characteristics including high confinement, long lifetime, strong field enhancement, broad spectral range, electrical tunability and a broad spectral range from terahertz to infrared [1-21]. So far, plasmons in single layer graphene (SLG) have been extensively studied and are generally well understood. One convenient way to create new plasmonic materials with novel physics and properties is by stacking graphene with graphene and other 2D materials into vdW materials or heterostructures. Indeed, the 2D nature of graphene makes it extremely sensitive to interlayer coupling that could dramatically modify the properties of Dirac fermions and their plasmonic excitations. For example, earlier studies about Bernal-stacked BLG [20,22] and graphene/hBN heterostructures [23,24] have demonstrated many unique plasmonic characteristics compared to those of SLG.

In this Letter, we report a systematic nano-infrared imaging study of plasmons in TBLG [Fig. 1(a)], which is formed when two misorientated graphene layers are stacked together by vdW forces. Depending on the twist angle ($\theta$) between the two graphene layers, moiré patterns with different periodicities could form [Fig. 1(b)]. Due to the interlayer coupling and modulation of Dirac fermions by moiré superlattice potential, the electronic structure of TBLG shows distinct features compared to SLG and Bernal-stacked BLG, and it varies systematically with $\theta$. For example, TBLG with a sizable $\theta$ features two separated Dirac cones [Fig. 1(c)] in the momentum space [25-34]. Moreover, the Fermi velocity ($v_F^{tBLG}$) close to the charge neutrality point is renormalized compared to that of SLG ($v_F^{SLG}$), namely $v_F^{tBLG}$ drops systematically below $v_F^{SLG}$ as $\theta$ decreases [25-30]. Therefore, TBLG is a unique system where the Fermi velocity of Dirac fermions could become an adjustable variable in experimental studies. The unique electronic properties of TBLG have led to observations of many interesting optical phenomena through far-field spectroscopic experiments [35-38]. So far, plasmonic responses of TBLG have not been explored experimentally despite the potential rich physics according to theoretical predictions [39,40].

Here we utilize a scattering-type scanning near-field optical microscope (s-SNOM) to perform nano-infrared imaging studies of TBLG plasmons. The s-SNOM apparatus is built based on an atomic force microscope (AFM). As illustrated in Fig. 1(a), the infrared light (solid arrow) from a continuous-wave infrared laser is focused at the apex of a metalized AFM tip. The laser-illuminated tip acts as both a launcher and a detector of surface plasmons [13-23]. The back-scattered light (dashed arrow) off the tip-sample system contains essential information about plasmons underneath the tip. The s-SNOM collects simultaneously the topography, near-field scattering amplitude (*s*) and phase ($\psi$). By analyzing both the *s* and $\psi$ data images, we can determine the key plasmonic parameters of TBLG. Our samples were grown by the chemical vapor deposition (CVD) method on copper foils [41-43] and then transferred to the standard $SiO_2$/Si substrates (Supplemental Material [44]). As shown in Fig. 1(d), both SLG and TBLG are single-crystal grains with a hexagonal shape and the TBLG grains are typically located at the center of relatively larger SLG grains. Occasionally, we also see hexagon-like shapes with slightly curved edges (Fig. S5) [45,46], but in all cases, these SLG or TBLG single-crystals demonstrate a six-fold rotational symmetry (Supplemental Material [44]). According to the previous studies [45,47], the six-fold flake symmetry correlates strictly and accurately with the lattice orientation, so it is convenient to determine the twist angle with a relatively good accuracy (±1°) by comparing the orientations of the TBLG and SLG grains.

Representative s-SNOM imaging data are shown in Fig. 2, where we plot both the normalized amplitude [Fig. 2(b)] and phase [Fig. 2(c)] signals of a typical sample region containing two TBLG grains. The data images were taken at an excitation laser energy of $E = 0.11$ eV that is away from the strong optical phonon resonance of $SiO_2$ centered at about 0.14 eV [13]. Therefore, the near-field responses of graphene at our excitation energy are mainly due to plasmons [14]. Figure 2(a) sketches the sample configuration, where we can conveniently determine the twist angles of TBLG from the orientations of the hexagonal grains. For example, the TBLG sample labeled as 'TBLG1' has a twist angle of about 26° relative to SLG and the one labeled as 'TBLG2' has a twist angle close to 1°. Their different twist angles result in distinct near-field responses. As shown in Figs. 2(b) and 2(c), TBLG1 has significant higher near-field amplitude compared to SLG but shows no clear phase contrast with respect to the latter. On the contrary, the amplitude of TBLG2 is almost the same as that of SLG and its phase is slightly weaker. Such dramatic differences in the near-field responses are clear indications of the strong $\theta$ dependence of TBLG. More near-field data images are given in Figs. S1 and S2, where additional TBLG samples with various twist angles are shown. In all the samples we measured (partly shown in Figs. 2, S1 and S2), the near-field amplitude of TBLG is comparable to SLG for $\theta \leq 3°$, and gradually increases from an intermediate signal ($\theta \approx 5°$) to a maximum value ($\theta > 7°$). The phase signal of TBLG, on the other hand, is roughly the same as SLG for $\theta > 7°$, but slightly declines as $\theta$ approaches 0°. The above $\theta$ dependence is more clearly seen in Figs. 4(c) and 4(d), where we summarized the extracted amplitude and phase signal data points (squares) from tens of TBLG samples that we measured.

The unique near-field responses discussed above are directly linked to the plasmons in TBLG. Indeed, we found direct evidence of plasmons in the high-resolution imaging data (Figs. 3 and S3) taken over five small sample regions (marked with dashed squares) in Fig. S1. These regions (labeled with 'P1' – 'P5') are chosen to be at the edge of SLG or the boundaries between SLG and TBLG. The amplitude images are shown in Figs. 3(a) – 3(e), where we observe bright fringe(s) close to the SLG edge and the SLG/TBLG boundaries. This can be seen more clearly in the line profiles [grey solid curves in Fig. 3(f) – 3(j)] taken perpendicular to the edges or boundaries in the amplitude images (along blue dashed lines). Here in these line profiles, the peak features correspond to the bright fringes in the images.

According to previous studies [14-24], the bright fringes registered by the s-SNOM are generated due to the constructive interference between tip-launched and edge- or boundary-reflected plasmons. The plasmonic origin of the observed fringes is further confirmed by the spectroscopic imaging data (Fig. S4), where we observed a systematic evolution of the bright fringes with laser energy, consistent with the dispersion nature of plasmons. There are two main observations from these plasmonic fringes data (Fig. 3). First, fringes are clear and strong close to the SLG/TBLG3 ($\theta \approx 27°$) and SLG/TBLG4 ($\theta \approx 12°$) boundaries. As $\theta$ decreases, the fringes become weaker and fewer at the SLG/TBLG5 ($\theta \approx 5°$) boundary and then barely seen at the SLG/TBLG6 boundary ($\theta \approx 3°$). Second, in the case of SLG/TBLG3 [Fig. 3(b)] and SLG/TBLG4 [Fig. 3(c)] boundaries, we can easily identify two to three fringes. Nevertheless, at the edge of SLG [Fig. 3(a)], we can only see one bright fringe. Note that the edge of SLG is a nearly-perfect plasmon reflector. The plasmon reflection at the SLG/TBLG boundaries, on the other hand, is in principle weaker. Therefore, we can tell directly from the fringes data that our SLG sample has a relatively higher plasmon damping compared to TBLG with relatively large $\theta$. Figure S3 plot the near-field phase images and the corresponding line profiles, where plasmonic interference fringes are also seen. The amplitude and phase imaging data are consistent and complementary to each other. They are all considered in our modeling as discussed in detail below.

To extract quantitative information about plasmons in SLG and TBLG, we performed numerical modeling of both the plasmonic fringes profiles (Figs. 3 and S3) and the $\theta$-dependent near-field amplitude and phase signals [Figs. 4(c) and 4(d)] by using the so-called spheroid model. In this model, the s-SNOM tip is approximated as a highly-elongated conducting spheroid (Fig. S7) and we evaluate the complex scattering signal by computing the total radiating dipole of the coupled tip-sample system (Supplemental Material [44]). We wish to emphasize that our model has been proven to effective in describing s-SNOM responses of graphene with quantitative accuracy [14,16,23]. The main modeling parameter of the sample is the optical conductivity

($\sigma = \sigma_1 + i\sigma_2$) that is directly linked to the complex plasmon wavevector ($q_p = q_1 + iq_2$) under the long-wavelength approximation:

$$q_p \approx i\varepsilon_0(1+\varepsilon_s)E/\hbar\sigma. \quad (1)$$

Here $\hbar$ is the reduced plank constant, $\varepsilon_0$ is the vacuum permittivity, and $\varepsilon_s$ is the relative permittivity of $SiO_2$. For convenience, our analysis and discussions are based on the following two parameters: the plasmon wavelength ($\lambda_p = 2\pi/q_1$) and damping rate ($\gamma_p = q_2/q_1$). Based on Eq. 1, we know that the plasmon wavelength ($\lambda_p$) is roughly proportional to $\sigma_2$, and the damping rate ($\gamma_p$) scales linearly with $\sigma_1/\sigma_2$.

We first fit the plasmonic fringe profiles of SLG [Figs. 3(f) and S3(f)]. Through the fitting, we extract the plasmon wavelength ($\lambda_p^{SLG}$) and damping rate ($\gamma_p^{SLG}$) of SLG at $E = 0.11$ eV to be about 279 nm and 0.2, respectively. Based on Eq. (1), we can establish a simple relation between $\lambda_p^{SLG}$ and the carrier density ($n$) under the Drude approximation:

$$\lambda_p^{SLG} \approx \frac{2e^2\hbar v_F^{SLG}\sqrt{\pi|n|}}{(1+\varepsilon_s)\varepsilon_0 E^2}. \quad (2)$$

Here $v_F^{SLG} \approx 10^6$ m/s is the Fermi velocity of SLG. Equation (2) allows us to estimate the carrier density of SLG to be $n \approx 1.2 \times 10^{13}$ cm$^{-2}$, which is a typical value for highly-hole-doped CVD samples on $SiO_2$/Si substrates at ambient conditions [16]. The relatively high doping is mainly due to the impurities on the surface of $SiO_2$ as well as the water and oxygen molecules in the air [48]. Considering that all the samples studied here share the same substrate and atmospheric conditions, they are expected to share roughly an equal density of external dopants and therefore a similar carrier density [21,22].

Based on the extracted parameters of SLG, we then determine both the plasmon wavelength ($\lambda_p^{tBLG}$) and damping rate ($\gamma_p^{tBLG}$) of TBLG by fitting the fringe profiles at the SLG/TBLG boundaries (Figs. 3 and S3). Through fitting, we estimate that ($\lambda_p^{tBLG}$, $\gamma_p^{tBLG}$) of TBLG3 ($\theta \approx 27°$), TBLG4 ($\theta \approx 12°$), TBLG5 ($\theta \approx 5°$) and TBLG6 ($\theta \approx 3°$) to be (393 nm, 0.10), (387 nm, 0.11), (340 nm, 0.16) and (278 nm, 0.28), respectively. These numbers are plotted in Figs. 4(a) and 4(b) as data points. Note that the first two numbers of $\lambda_p^{tBLG}$ can be read out directly from the fringe profiles of TBLG3 and TBLG4 by doubling the fringe period [arrows in Figs. 3(g) and 3(h)]. Nevertheless, precise modeling of the complex fringe profiles is required to extract both $\lambda_p^{tBLG}$ and $\gamma_p^{tBLG}$, and to analyze data from TBLG samples without strong fringes [Figs. 3(d) and 3(e)]. In the latter case, the modeling fits mainly the $s$ and $\psi$ signals of TBLG in contrast to SLG. In Figs. 4(c) and 4(d), we show the modeling curve of $s$ and $\psi$ contrast signals of TBLG versus SLG at a wide distribution of twist angles (red curves), which match well the trend of the experimental data points with twist angles above 3° (marked with dashed lines). At twist angles below 3°, the experimental data points clearly deviate from the modeling curve, which will be discussed in the following paragraphs. The smooth $\lambda_p^{tBLG}(\theta)$ and $\gamma_p^{tBLG}(\theta)$ parameters [red curves in Figs. 4(a) and 4(b)] used to model the $s$ and $\psi$ contrast signals are fully consistent with the discrete data points obtained from fringe profile fitting.

Now we wish to discuss the origin of the $\theta$-dependence of $\lambda_p^{tBLG}$ and $\gamma_p^{tBLG}$. We first pay attention to twist angles above 3°, where TBLG is in the Dirac regime [Fig. 1(c)] [25-29]. Here we assume that carriers are equally distributed among the two graphene layers, which is reasonable considering no external gating. The general results won't change much even with slightly unequal carrier distribution among the two graphene layers (Fig. S6 and Supplemental Material [44]). Under the equal carrier distribution assumption, $\lambda_p^{tBLG}$ can be written as

$$\lambda_p^{tBLG}(\theta) \approx \frac{2e^2\hbar\sqrt{2\pi|n|}}{(1+\varepsilon_s)\varepsilon_0 E^2} v_F^{tBLG}(\theta), \tag{3}$$

where the Fermi velocity of TBLG ($v_F^{tBLG}$) is proven to be sensitively dependent on $\theta$ due to the Fermi velocity renormalization. The amount of Fermi velocity renormalization is determined by the interlayer coupling energy (*t*) of TBLG [inset of Fig. 4(a)] as described by the following equation: [25]

$$v_F^{tBLG}(\theta) = v_F^{SLG}[1 - 9(\frac{t}{\hbar v_F^{SLG} \Delta K})^2] . \tag{4}$$

Here, $\Delta K = (8\pi/3a)\sin(\theta/2)$ is the momentum separation of the two Dirac cones [Fig. 1(c)], and *a* = 0.246 nm is the lattice constant. Equation 4 indicates that *t* is the one single parameter that controls $v_F^{tBLG}(\theta)$ and hence $\lambda_p^{tBLG}(\theta)$. Here we set *t* to be 0.1 eV, which is roughly consistent with previous studies [29,34]. With such a *t* setting, we calculated $\lambda_p^{tBLG}(\theta)$ based on Eqs. (3) and (4), which is shown as the red curve in Fig. 4(a). Other choices of *t* will lead to either faster or slower decreasing of $v_F^{tBLG}$ and hence $\lambda_p^{tBLG}$ as $\theta$ drops [inset of Fig. 4(a)].

The origin for the higher $\gamma_p^{tBLG}$ at smaller $\theta$ in the Dirac regime ($\theta \geq 3°$) is likely due to the stronger charge scattering rates [49,50]. According to previous literature [51], the charge scattering rates ($\Gamma$) due to either long-range Coulomb scattering or short-range defect scattering are inverse proportional to the Fermi velocity. Therefore, as $\theta$ decreases, $\Gamma$ rises and thus $\gamma_p^{tBLG}$ increases. Note that interband transitions are forbidden due to the Pauli blocking for $\theta \geq 3°$, where threshold energy for interband transitions ($2E_F^{tBLG}$) is estimated to be over 0.2 eV, far above our laser energy (0.11 eV).

Finally, we wish to discuss briefly TBLG samples with twist angles below 3°, where the Dirac approximation begins to fail [26-29]. In this regime, we find that the amplitude signal of the TBLG samples deviates from the projected trend of the modeling curves and stay close to that of SLG [Fig. 4(c)]. With quantitative modeling (Fig. S8), we estimate that the $\lambda_p^{tBLG}$ at small twist angles ($\theta \leq 2°$) is in the range from 278 to 314 nm. According to previous theoretical studies [26-28], the lowest-energy bands of TBLG with small twist angles become flat or nearly-flat close to the charge neutrality point, which could lead to an extremely small $\lambda_p^{tBLG}$ (Eq. 3). The finite $\lambda_p^{tBLG}$ of TBLG ($\theta \leq 2°$) observed here suggests that the Fermi surface of our highly-doped samples is away from these relatively flat bands. The phase signals [Fig. 4(d)] of TBLG ($\theta \leq 2°$) appear to be slightly smaller than that of SLG and large-twist-angle TBLG, indicating even higher plasmon damping rates: $\gamma_p^{tBLG}$ ($\theta \leq 2°$) = 0.2 - 0.4 (Fig. S8). The higher damping is most likely due to interband transitions, which are enabled in TBLG ($\theta \leq 2°$) at our excitation energy (0.11 eV) due to the small energy separations between the lowest-energy bands. More detailed discussions about TBLG with $\theta \leq 2°$ are given in the Supplemental Material [44]. Future studies with more comprehensive experiments of small-twist-angle TBLG and more precise determinations of twist angles are needed to explore further TBLG plasmons in the non-Dirac regime.

By combining the state-of-the-art s-SNOM technique with rigorous numerical modeling, we performed a systematic nano-infrared imaging study of TBLG single crystals with various twist angles. In the Dirac linear regime, we found that TBLG support infrared plasmons with parameters that vary systematically with the twist angle between the two graphene planes. The underlining physics behind the observed twist angle dependence is the Fermi velocity renormalization, which is originated from the interlayer electronic coupling. Our study establishes TBLG as a unique platform where the Fermi velocity, the fundamentally important parameter of Dirac fermions, has become an adjustable variable in nano-optical and plasmonic studies of Dirac materials.

F.H., Y.L. and Z.F. acknowledge startup support from Iowa State University and the royalty funds from the U.S. Department of Energy, Ames Laboratory. The nano-optical setup is partially supported by W. M. Keck Foundation. S.R.D. acknowledges the U.S. Department of Energy, Ames Laboratory. The synthesis of TBLG samples at Purdue University is partially supported by NSF CMMI (grant 1538360) and NSF EFMA (Grant No. 1641101).

**Figure Captions**

FIG. 1. (a) Illustration of the nano-infrared imaging experiment of a TBLG/SLG single crystal. The solid and dashed arrows mark the directions of the incident laser beam and back-scattered photons, respectively. (b) Sketch of the crystal structure of TBLG revealing moiré periodic pattern. The double-sided arrow marks the moiré period. (c) Calculated band structure of TBLG with a twist angle of 5° with the continuum model [25]. Here the momentum unit $\Delta K$ equals to the separation between the two Dirac points ($K_1$ and $K_2$) in the momentum space. (d) Optical image of representative TBLG/SLG single-crystal samples. The scale bar represents 5 μm.

FIG. 2. (a) Sketch of the sample geometry indicating two adjacent TBLG grains with different twist angles with respect to SLG. (b) and (c) The near-field images plotting scattering amplitude ($s$) and phase ($\psi$), respectively. In both images, the amplitude or phase signal is normalized to that of SLG. The laser energy is set to be at $E = 0.11$ eV. The scale bars represent 3 μm.

FIG. 3. (a)-(e) High-resolution near-field amplitude images of the five small regions ('P1' – 'P5') marked in Fig. S1 (squares), respectively. The white dashed lines in the images mark the SLG edge and the TBLG/SLG boundaries. The scale bars represent 400 nm. (f)-(j), Experimental (grey solid)

and modeled (red dashed) amplitude profiles taken along the blue dashed lines in the corresponding near-field images above. The blue arrows in (g) and (h) mark the size of $\lambda_p^{tBLG}$ that is twice the fringe period. The vertical dashed lines mark the boundaries between SLG and TBLG.

FIG. 4. (a) The $\lambda_p^{tBLG}$ data points extracted by modeling the fringe profiles in Fig. 3. Inset plots the calculated $v_F^{tBLG}(\theta)$ normalized to $v_F^{SLG}$ considering different $t$. (b) The $\gamma_p^{tBLG}$ data points extracted by modeling the fringe profiles in Fig. 3. The red curves in (a) and (b) are used for calculations of the amplitude and phase signals in (c) and (d). The blue arrows in (a) and (b) mark the values of $\lambda_p^{SLG}$ and $\gamma_p^{SLG}$, respectively. (c) The $\theta$-dependent near-field amplitude from both experiment (squares) and modeling (red curve). (d) The $\theta$-dependent near-field phase from both experiment (squares) and modeling (red curve). Both the amplitude (c) and phase (d) of TBLG are normalized to those of SLG. The vertical dashed lines in (c) and (d) mark $\theta = 3°$.

**Figure 1**

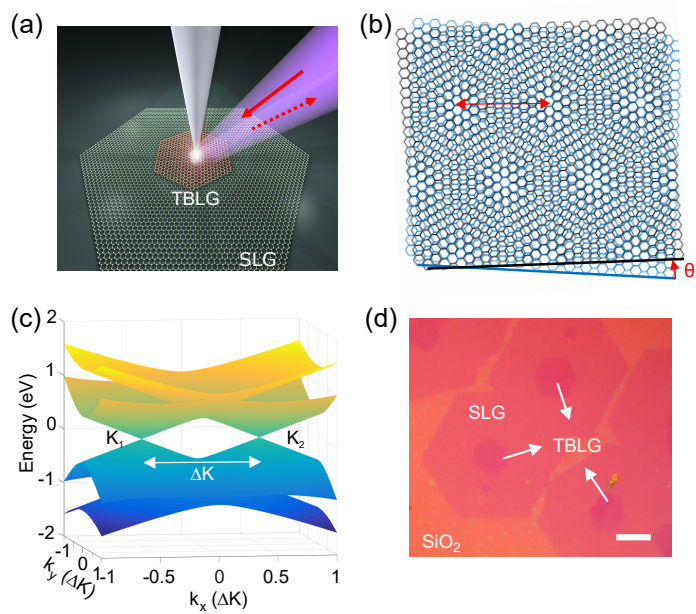

**Figure 2**

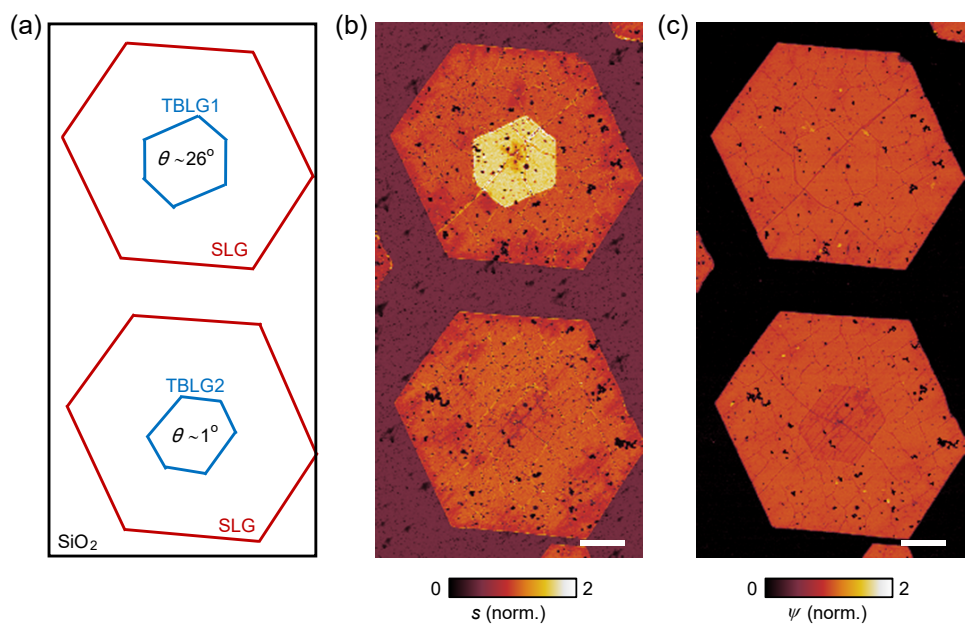

**Figure 3**

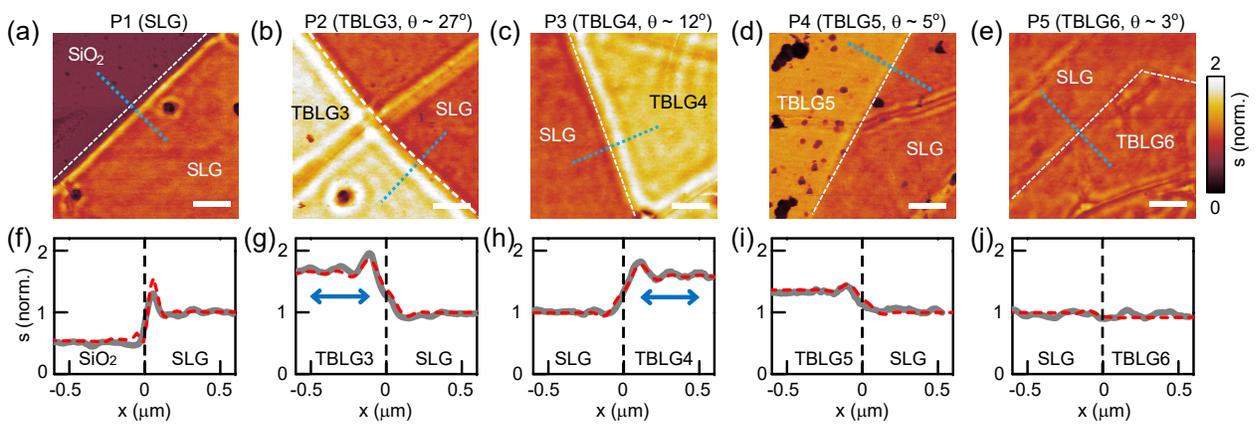

**Figure 4**

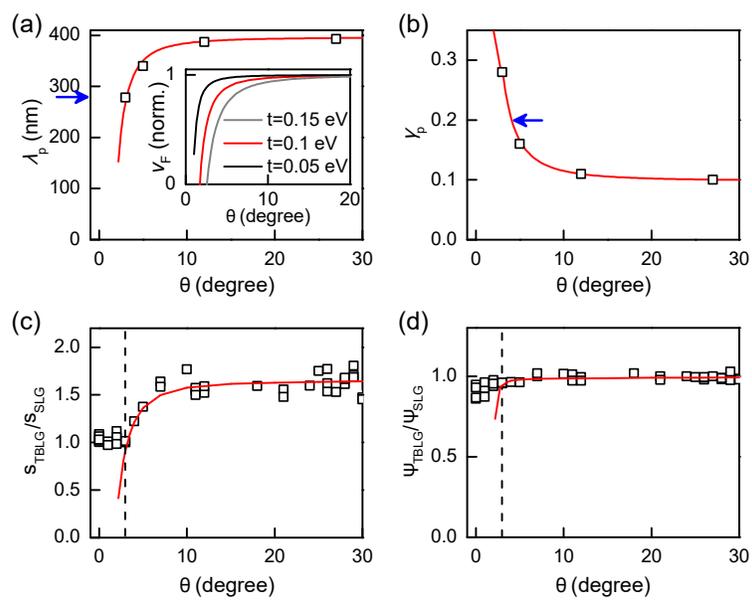

# Supplemental Material
# Real-Space Imaging of the Tailored Plasmons in Twisted Bilayer Graphene

F. Hu[1,2]*, Suprem R. Das[2,3,4,5]*, Y. Luan[1,2]*, T.-F. Chung[6,7], Y. P. Chen[6,7,8,9], Z. Fei[1,2]†

[1]Department of Physics and Astronomy, Iowa State University, Ames, Iowa 50011, USA
[2]Ames Laboratory, U.S. Department of Energy, Iowa State University, Ames, Iowa 50011, USA
[3]Department of Mechanical Engineering, Iowa State University, Ames, Iowa 50011, USA
[4]Department of Industrial and Manufacturing Systems Engineering, Kansas State University, Manhattan, KS 66506, USA
[5]Department of Electrical and Computer Engineering, Kansas State University, Manhattan, KS 66506, USA
[6]Birck Nanotechnology Center, Purdue University, West Lafayette, Indiana 47907, USA
[7]Department of Physics and Astronomy, Purdue University, West Lafayette, Indiana 47907, USA
[8]School of Electrical and Computer Engineering, Purdue University, West Lafayette, Indiana 47907, USA
[9]Purdue Quantum Center, Purdue University, West Lafayette, Indiana 47907, USA

*These authors contributed equally to this work.

† Email: (Z.F.) zfei@iastate.edu

**List of contents**

1. Nano-optical imaging setup

2. Sample fabrication procedures

3. Additional s-SNOM imaging data

4. Flake shape of single-crystal SLG/TBLG samples

5. Unintentional doping of the samples

6. Analysis of unequal carrier distribution

7. Quantitative modeling of the s-SNOM signals

8. Analysis and discussions of small-twist-angle TBLG

## 1. Nano-infrared imaging setup

For nano-infrared imaging experiments in the current work, we used a scattering-type scanning near-field optical microscope (s-SNOM) apparatus (www.neaspec.com) that is built based on an atomic force microscope (AFM). The AFM is operated in the tapping mode with a tapping frequency of about 270 kHz and a tapping amplitude of about 50 nm. The AFM tips used in the current work are platinum iridium coated silicon tips with a radius of curvature of about 25 nm at the tip apex (www.nanoworld.com). For signal detection, we used a mercury cadmium telluride photodiode (www.kolmartech.com). A pseudo-heterodyne interferometer is implemented in our s-SNOM to extract both the near-field amplitude ($s$) and phase ($\psi$) of the complex near-field signal. To suppress the background signal, we demodulated the near-field signal at the $3^{rd}$ harmonics of the tapping frequency of the AFM tip. For optical excitation, we used a continuous-wave $CO_2$ laser (www.accesslaser.com). The photon energy of the laser can be discretely tunable from 0.11 to 0.13 eV. Our nano-infrared imaging experiments were all performed at ambient conditions.

## 2. Sample fabrication procedures

The growth of the hexagon-shaped TBLG/SLG single crystals was achieved by using the atmospheric-pressure chemical vapor deposition (CVD) method. A mixture of methane carrier gas was atomically cracked at high temperature (1050 °C) with argon/hydrogen gas before controllably getting deposited onto a pre-cleaned copper foil. After the growth process, the methane flow was stopped, and the sample was cooled down to room temperature in the furnace with argon/hydrogen gas flow uninterrupted. For s-SNOM studies, our samples on copper foil were transferred onto the standard $SiO_2$/Si wafers used a wet-chemical etching and transfer method. In short, a 100-nm-thick polymethyl methacrylate (PMMA) protection layer was applied to one side of the copper foil and the TBLG on the other side of the foil was plasma etched using an oxygen plasma. The copper was then etched overnight in an etchant solution. After multiple washing of the floating TBLG/PMMA stack with deionized water and a mild aqueous acid, the stack was transferred onto the $SiO_2$/Si wafers. The PMMA layer was stripped off TBLG in about 4 hours after baking/drying on a hotplate at 90 °C. Finally, the sample was electronically cleaned at 300 °C for 2 hours. More detailed introductions about the growth procedures are given in the earlier study (Ref. [41] in the main text).

## 3. Additional s-SNOM imaging data

In Figs. S1-S4, we provide additional near-field imaging data about TBLG. Figure S1 plots four typical TBLG crystals ('TBLG3' – 'TBLG6') that we used for high-resolution imaging. The high-resolution imaging data taken at five specific locations ('P1' to 'P5' marked in Fig. S1) are plotted in Fig. 3 (amplitude images and profiles) of the main text and Fig. S3 (phase images and profiles). Figure S2 plots near-field amplitude and phase images at four different locations, where a total of eight SLG/TBLG crystals ('TBLG7' – 'TBLG14') are included. Figure S4 plots the excitation-energy-dependent s-SNOM imaging data at the boundary between 'SLG' and 'TBLG3'. One can see that the plasmonic interference fringes at the SLG/TBLG3 boundary show a systematic variation with photon energy. Through modeling of the fringe profiles, we extracted the plasmon wavelength of TBLG (SLG) to be about 393 nm (279 nm), 370 nm (262 nm) and 350 nm (248 nm) for excitation energies of 0.110, 0.116 and 0.121 eV, respectively. The energy-dependent plasmon wavelengths are consistent with the dispersion nature of plasmons.

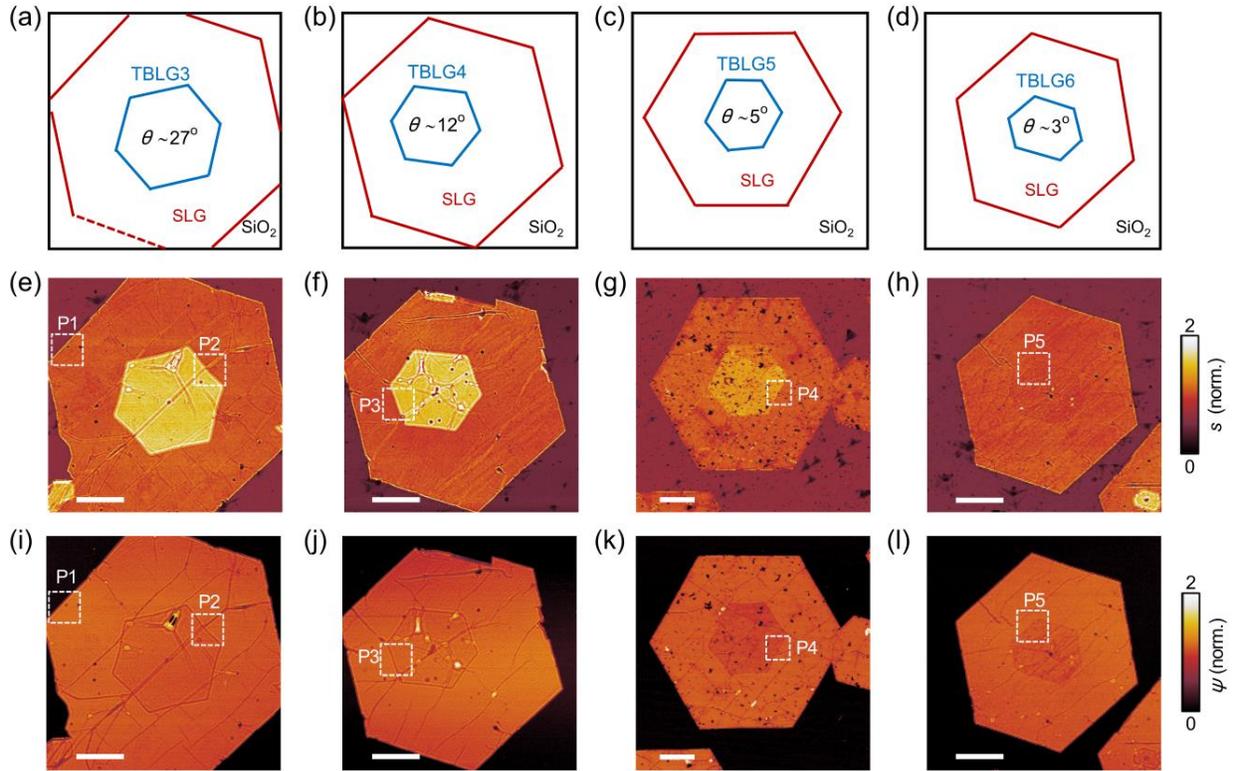

FIG. S1. (a)-(d) Sketches of the sample geometries of four additional TBLG single crystals ('TBLG3' – 'TBLG6') with different twist angles. (e)-(h) The near-field amplitude images of the samples sketched in (a)-(d), respectively. (i)-(l) The near-field phase images of the samples sketched in (a)-(d), respectively. The excitation laser energy is set to be at $E = 0.11$ eV. In all the images, the amplitude or phase signal is normalized to that of SLG. The squares mark the five positions ('P1' ─ 'P5') where we collected high-resolution imaging data as shown in Fig. 3 of the main text and Fig. S3 of the Supplemental Material. The scale bars represent 3 μm.

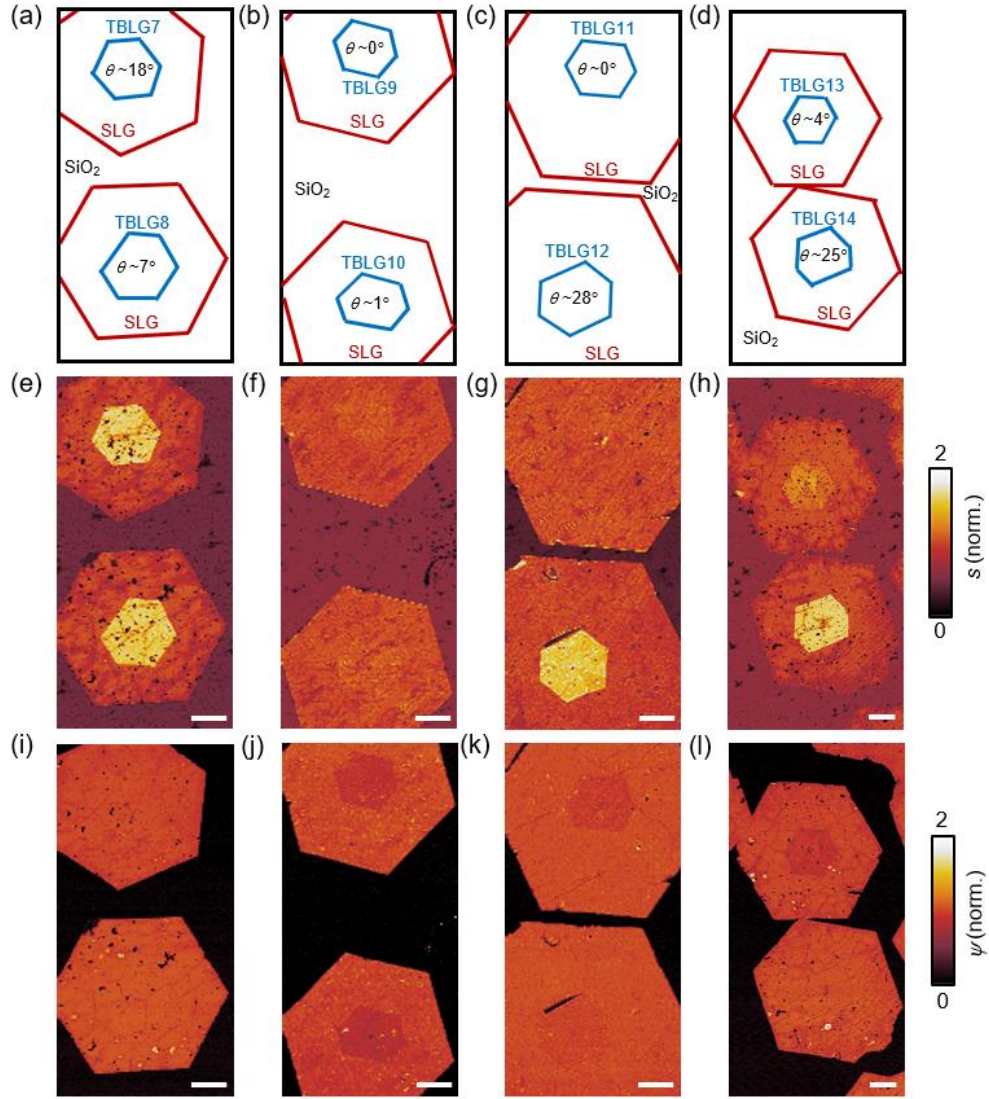

FIG. S2. (a)-(d) Sketches of the sample geometry of four samples regions that include a total of eight additional TBLG single crystals ('TBLG7' – 'TBLG14'). (e)-(h) Near-field amplitude images of sample regions sketched in (a)-(d), respectively. (i)-(l) The near-field phase images of sample regions sketched in (a)-(d), respectively. In all the images, the amplitude or phase signal is normalized to that of SLG. The scale bars represent 3 μm.

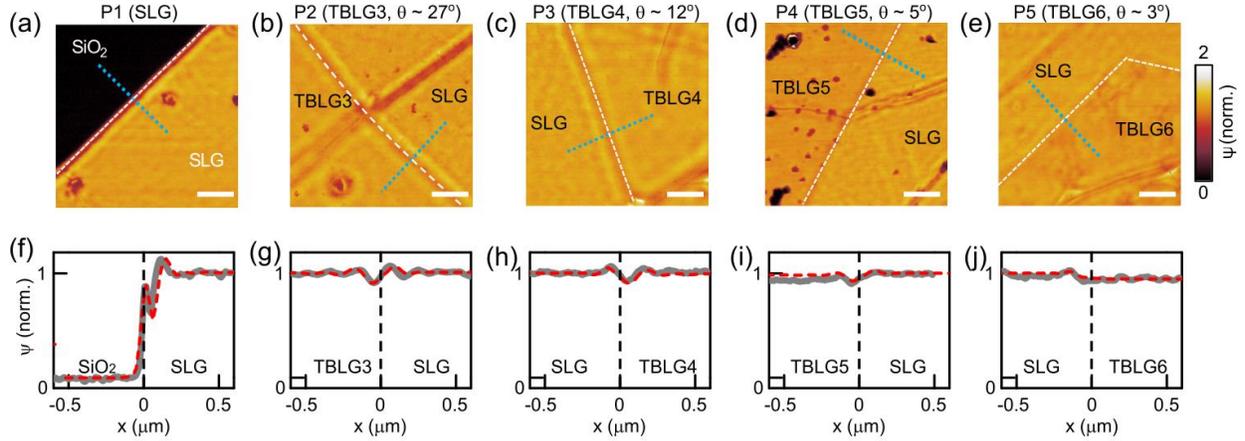

FIG S3. (a)-(e) Near-field phase images of the five small regions ('P1' – 'P5') marked in Fig. S1 (squares). The white dashed lines in the images mark the edges of SLG and the boundaries between TBLG and SLG. The scale bars in all the images represent 400 nm. (f)-(j) Experimental (grey solid) and modeled (red dashed) phase profiles taken perpendicular to the SLG edge and the TBLG/SLG boundaries. In all the near-field images, the phase signal is normalized to that of SLG. The experimental phase profiles were taken along the blue dotted lines in the corresponding near-field images. The vertical dashed lines mark the boundaries between SLG and TBLG.

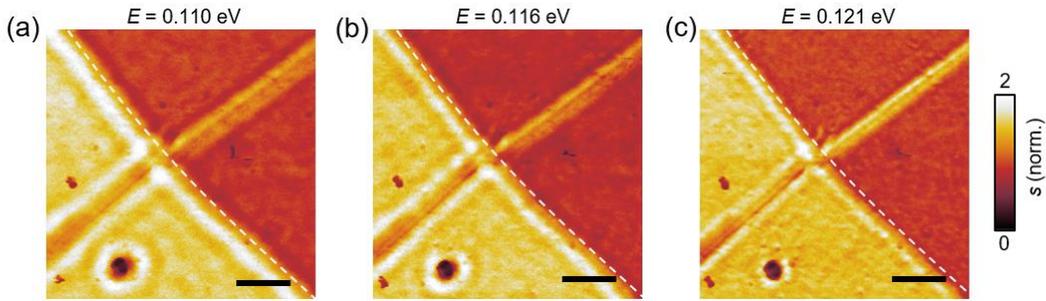

FIG. S4. Near-field amplitude images of the SLG/TBLG3 boundary (see Fig. 3 in the main text) at various excitation laser energies: $E = 0.110$ eV (a), 0.116 eV (b) and 0.121 eV (c). Here we plot the near-field amplitude normalized to that of SLG. The scale bars represent 400 nm.

## 4. Flake shape of single-crystal SLG/TBLG samples

Most of our single-crystal TBLG/SLG samples appear in a hexagonal shape with straight edges. Occasionally, we also see hexagon-like shapes with slightly-curved edges. For example, in Fig. S5a (or Fig. 3b), we found the edge of TBLG3 is not exactly straight. A large-area image is shown in Fig. S5b, where we mark carefully the boundaries of SLG and TBLG3 grains with green and blue curves. These curves are replotted in Fig. S5c, where one can see that the edges of the TBLG3 grain are not straight. Instead, it has a hexagonlike shape with slight negative curvature at the edges. This is, in fact, one of the typical shapes of single-crystal graphene flakes. As reported by an earlier study (Ref. [45] in the main text), various types of six-fold symmetric shapes of graphene flakes (from flower-like to hexagon-like) could occur by varying systematically the CVD growth conditions, among which both the perfect hexagonal shape and slight-curved hexagonal shape are included. The lattice orientation of graphene is consistent with the six-fold symmetry of the flake, as verified by diffraction experiments in Refs. [45] and [47] of the main text. An earlier study also shows slightly-curved hexagon shape in twisted bilayer and trilayer graphene flakes

(Ref. [46] in the main text). To sum up, the TBLG3 sample with slight-curved edges is also typical for CVD-grown single-crystal samples and we followed the symmetry of the entire flake when determining the twist angle.

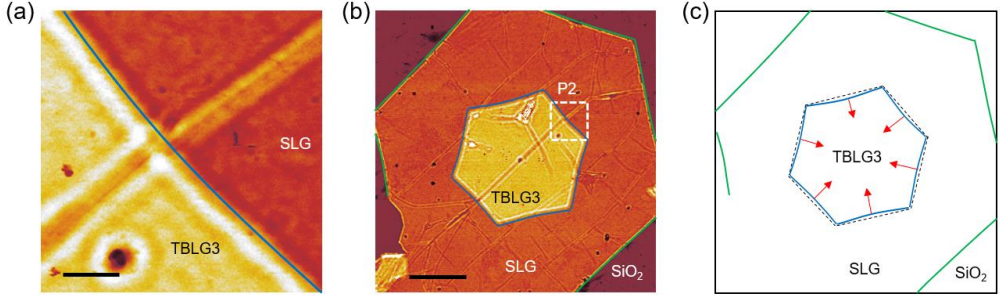

FIG. S5. (a) High-resolution near-field amplitude image of the TBLG3/SLG boundary, a replot of Fig. 3b in the main text. (b) The large-area near-field amplitude of the TBLG3/SLG sample, a replot of Fig. S1e. (c) Sketch of the geometry of the TBLG3/SLG sample.

## 5. Unintentional doping of the samples

Our CVD-grown samples are all highly hole-doped at ambient conditions due to the high density of impurities on the amorphous $SiO_2$ surface as well as water and oxygen molecules in the air (Ref. [48] in the main text). Considering that all the data presented in the paper were from samples produced from the same batch and transferred onto the same wafer, they share nearly identical chemical and environmental conditions. Therefore, we are safe to assume that all the TBLG and SLG samples share roughly an equal density of external dopants and therefore a similar carrier density. Indeed, previous studies have confirmed that adjacent SLG and Bernal-stacked or randomly-stacked BLG share roughly the same carrier density (Refs. [21,22] in the main text). By fitting the plasmonic fringe profile of SLG, we can accurately determine the plasmon wavelength of SLG be about 279 nm, based on which we estimate the carrier density of SLG and TBLG is $n \approx 1.2 \times 10^{13}$ cm$^{-2}$.

## 6. Analysis of unequal carrier distribution

In the current paper, we assume equal carrier distribution among the top and bottom layers of TBLG when analyzing the data. This is a reasonable assumption considering the following facts. First, there is no gating or dedicated chemical doping to introduce doping asymmetry, so strong asymmetry is unlikely to occur due to the natural doping by the air or substrate. Second, even with small doping asymmetry, the results won't change much from the current analysis based on equal carrier distribution. Detailed analysis and discussions about doping asymmetry are given below.

Introducing unequal carrier distribution among the two graphene layers (or equivalently, adding a built-in electric field between the two graphene layers) can indeed affect the optical conductivity and plasmon wavelength of TBLG, and hence the near-field scattering signals. Here we only consider the Dirac linear regime (twist angle $\theta \geq 3°$), where the plasmon wavelength of TBLG obeys the following relationship:

$$\lambda_p^{tBLG} \propto \sigma_2^{tBLG} = \sigma_2^{top} + \sigma_2^{bottom} \propto E_F^{top} + E_F^{bottom} \propto v_F^{tBLG}(\sqrt{n^{top}} + \sqrt{n^{bottom}}), \quad (1)$$

where the superscripts "top" and "bottom" denote the top and bottom graphene layers, respectively. The total carrier density $n^{total} = n^{top} + n^{bottom}$. For convenience, we define the parameter $x = (n^{top} - n^{bottom})/n^{total}$ to characterize the doping asymmetry, which is in the range [-1, 1]. We do not consider $x < -1$ or $x > 1$, which means opposite charges distributed among the two graphene layers – an unlikely situation without dual gating setup. With the parameter $x$, Eq. 1 can be re-written as

$$\lambda_p^{tBLG} \propto v_F^{tBLG}(\sqrt{1+x}+\sqrt{1-x})\sqrt{n^{total}/2}. \qquad (2)$$

Accordingly, we can obtain the formula of SLG assuming the same total carrier density: $\lambda_p^{SLG} \propto v_F^{SLG}\sqrt{n^{total}}$, so the ratio between $\lambda_p^{tBLG}$ and $\lambda_p^{SLG}$ can be written as

$$\lambda_p^{tBLG}/\lambda_p^{SLG} = (v_F^{tBLG}/v_F^{SLG})(\sqrt{1+x}+\sqrt{1-x})/\sqrt{2}. \qquad (3)$$

In the Dirac regime of TBLG ($\theta \geq 3°$), $v_F^{tBLG}/v_F^{SLG} = \eta$ is the Fermi velocity renormalization factor that is a constant at every given twist angle, so $\lambda_p^{tBLG}/(\eta\lambda_p^{SLG}) = (\sqrt{1+x}+\sqrt{1-x})/\sqrt{2}$ is solely dependent on $x$. In Fig. S6, we plot the $x$-dependence curve of $\lambda_p^{tBLG}$ normalized to $\eta\lambda_p^{SLG}$, namely $\lambda_p^{tBLG}/(\eta\lambda_p^{SLG})$. Here $\lambda_p^{tBLG}/(\eta\lambda_p^{SLG})$ reaches the maximum $\sqrt{2}$ when $x = 0$ (namely equal carrier distribution among graphene layers). Introducing doping asymmetry ($x \neq 0$) will reduce the plasmon wavelength of TBLG. The $\lambda_p^{tBLG}/(\eta\lambda_p^{SLG})$ reaches the minimum 1 when $x = -1$ or 1 (carriers are fully distributed in the top or bottom graphene layer). From Fig. S6, we also notice that the variation of $\lambda_p^{tBLG}/(\eta\lambda_p^{SLG})$ is very small when $x$ is in a reasonable range. For example, for $|x| \leq 0.5$, the variation of $\lambda_p^{tBLG}/(\eta\lambda_p^{SLG})$ is only within 3% (see Fig. S6). In the case of our highly-doped TBLG samples ($n^{total} \approx 1.2 \times 10^{13}$ cm$^{-2}$), $|x| = 0.5$ means that the one layer has a carrier density of $0.3 \times 10^{13}$ cm$^{-2}$ corresponding to a Fermi energy of 0.2 eV, and the other layer has $0.9 \times 10^{13}$ cm$^{-2}$ corresponding to a Fermi energy of 0.35 eV. This is, in fact, a huge doping asymmetry that in principle cannot be created without external gating.

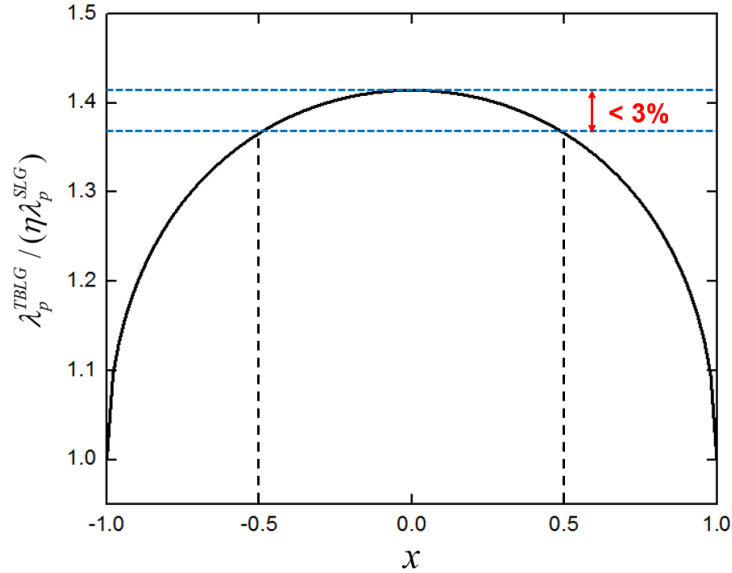

FIG. S6. The asymmetric doping dependence of the $\lambda_p^{TBLG}/(\eta\lambda_p^{SLG})$. The parameter $x = (n^{top} - n^{bottom})/n^{total}$ characterizes the doping asymmetry.

## 7. Quantitative modeling of the s-SNOM signals

To model the near-field amplitude and phase signals as well as the plasmon fringe profiles, we model our AFM tip as a metallic spheroid (Refs. [14,16,23] in the main text): the length of the spheroid is $2L$ and the radius of curvature at the tip ends is $a$ (Fig. S7). Here, $a$ is set to be about 25 nm according to the manufacturer and $L$ is set to be 500 nm. Note that $L$ is not a very sensitive parameter in the modeling when $L \gg a$. The complex near-field signal (before demodulation) scales with the total radiating dipole $\mathbf{p}$ of the spheroid. Therefore, to compute the near-field signal, we calculate $\mathbf{p}$ at different $z$ coordinates of the lower end of the AFM tip. By calculating $\mathbf{p}$ at different $z$, we can perform 'demodulation' to get the $n^{\text{th}}$ harmonics of the near-field signal ($n = 3$ in this work). To calculate the line profiles perpendicular to the fringes, we also considered the in-plane coordinate ($x$) of the tip. Calculating $\mathbf{p}$ at different $x$ and $z$ coordinates allows us to plot the modeling profiles of near-field amplitude and phase. The modeling parameters of the sample (SLG or TBLG) are ($\sigma_1$, $\sigma_2$), or more conveniently ($\lambda_p$, $\gamma_p$). As introduced in the main text, $\lambda_p$ is roughly proportional to $\sigma_2$, and $\gamma_p$ scales linearly with $\sigma_1/\sigma_2$.

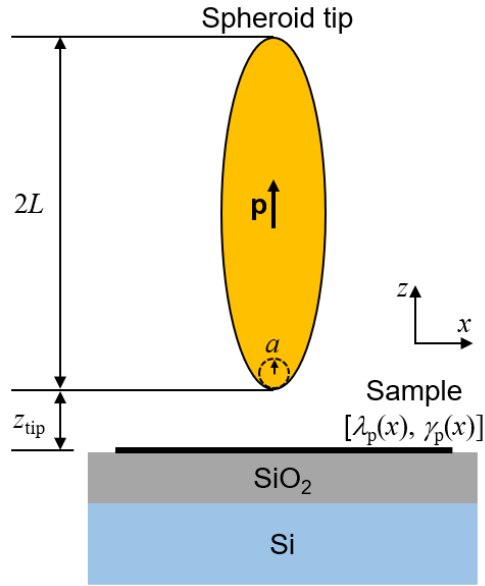

FIG. S7. Sketch of the spheroid model that we used to model the plasmonic responses of SLG and TBLG.

## 8. Analysis and discussions of small-twist-angle TBLG

In the main text, our analysis and discussions are mainly focused on TBLG samples with relatively large twist angles ($\theta \geq 3°$), where Dirac dispersion approximation stays true. Here, we wish to add more discussions about plasmonic responses of TBLG samples with small twist angles ($\theta \leq 2°$) based on our experimental data.

(1) Near-field amplitude and phase of TBLG ($\theta \leq 2°$).

For convenience, we first summarize in Fig. S8 the near-field signal (s, $\psi$) data points of all measured TBLG samples with small twist angles ($\theta \leq 2°$, red dots). From Fig. S8, one can see that the amplitude ratio between TBLG ($\theta \leq 2°$) and SLG ($s_{\text{tBLG}} / s_{\text{SLG}}$) is $1.04 \pm 0.08$ and the phase ratio between TBLG ($\theta \leq 2°$) and SLG ($\psi_{\text{tBLG}} / \psi_{\text{SLG}}$) is $0.92 \pm 0.07$. In other words, the amplitude signal of TBLG ($\theta \leq 2°$) is close to that of SLG with small variations, and the phase signal of TBLG ($\theta \leq 2°$) is slightly weaker than that of SLG with small variations.

(2) Plasmon wavelength and damping rate of TBLG ($\theta \leq 2°$).

As discussed in the main text, the near-field signals ($s$ and $\psi$) are directly linked to the plasmonic parameters ($\lambda_p$ and $\gamma_p$) of SLG and TBLG. Based on the $s$ and $\psi$ experimental data points (Fig. S8), we estimate with numerical modeling that the plasmon wavelength ($\lambda_p$) of TBLG ($\theta \leq 2°$) is in the range of 278 to 314 nm and the plasmon damping rate ($\gamma_p$) of TBLG ($\theta \leq 2°$) is in the range of 0.2 to 0.4. In Fig. S8, we also mark SLG (green star) and large-twist-angle TBLG ($\theta \rightarrow 30°$, blue shaded region). Compared to SLG, small-twist-angle TBLG ($\theta \leq 2°$) has higher $\gamma_p$ and the roughly equal $\lambda_p$ (or slightly higher, but less than 13% variation). From Fig. S8, one can also see that the smaller scattering phase of TBLG ($\theta \leq 2°$) compared to SLG is mainly due to the *overall higher $\gamma_p$*. The lack of amplitude contrast between TBLG ($\theta \leq 2°$) and SLG is mainly due to their *roughly equal $\lambda_p$*.

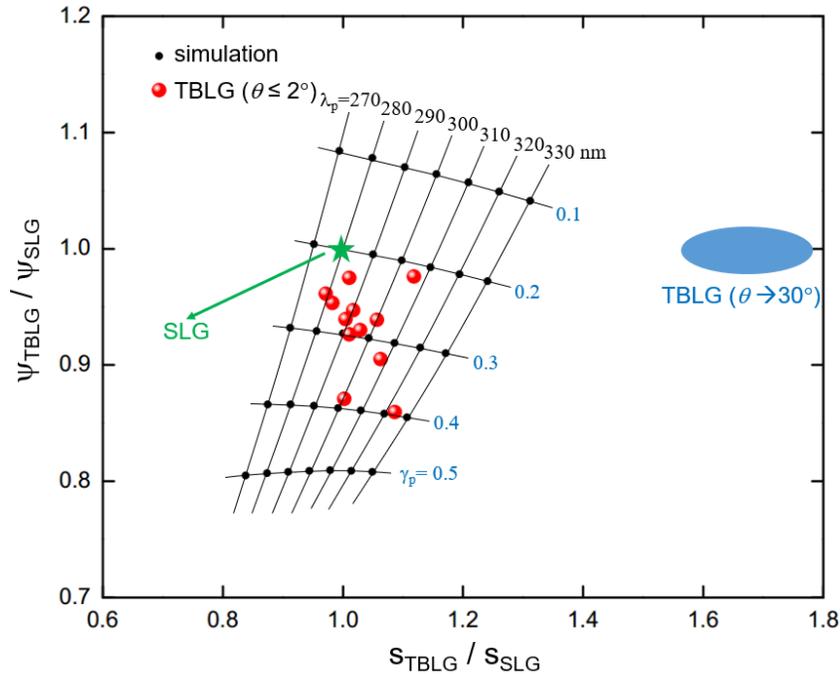

FIG. S8. Statistics analysis of scattering amplitude versus scattering phase signals of TBLG, both of which are normalized to those of SLG. Red dots summarize experimental data points of TBLG with $\theta \leq 2°$. Black dots are theoretical calculations assuming various ($\lambda_p$, $\gamma_p$) settings. Black curves connecting the black dots are drew to guide the eye. Green star and blue shaded region mark SLG and large-angle TBLG ($\theta \rightarrow 30°$) respectively.

(3) Further discussions of TBLG ($\theta \leq 2°$).

Based on the analysis and modeling shown above, we can conclude that all the small-twist-angle TBLG samples ($\theta \leq 2°$) we measured have a *roughly equal $\lambda_p$* and an *overall higher $\gamma_p$* compared to those of SLG. Now we wish to discuss these interesting findings. The *roughly equal $\lambda_p$* of TBLG ($\theta \leq 2°$) compared to SLG is particularly surprising at first glance considering that TBLG could become flat or nearly-flat bands as $\theta$ approaches 0°. A relatively flat band generally means an extremely-small plasmon wavelength if the Fermi surface lies on the band. Nevertheless, according to previous first-principle calculated band structures (Ref. [26] in the main text), the bands of TBLG at small twist angles becomes relatively flat only close to the charge neutrality point. The Fermi energy of our highly hole-doped TBLG samples is in principle far away from the charge neutrality point, where the band at the Fermi surface stay dispersive even for small-twist angle TBLG. For example, estimation based on Fig. 2 in Ref. [26] of the main text indicates that the Fermi surface of TBLG with $\theta = 2°$ is about -0.03 eV, where the band stays dispersive. Estimations based on Fig. 3 in Ref. [26] of the main text suggest that highest-energy valence band of TBLG (where the bands are relatively flat) with $\theta = 1.6°$, $1.3°$ and $1.0°$ is completely empty and the Fermi surface moves to lower valence bands, where the bands stay dispersive.

The *overall higher* $\gamma_p$ of TBLG ($\theta \leq 2°$) compared to SLG is most likely due to Landau damping caused by interband transitions. While interband transitions at our excitation laser energy (0.11 eV) are forbidden in SLG or TBLG with relatively large $\theta$. For $\theta \leq 2°$, interband transitions are enabled due to the small energy separations between lowest-energy bands. For example, according to the calculated band structures in Fig. 2 and Fig. 3 in Ref [26] of the main text, interband transitions at our excitation laser energy (0.11 eV) are forbidden in TBLG with $\theta = 2.9°$, but are fully enabled for $\theta = 2.0°$, $1.9°$, $1.6°$, $1.3°$ and $1.0°$.